\documentclass[letterpaper, 10 pt, conference]{ieeeconf}  % Comment this line out if you need a4paper

\IEEEoverridecommandlockouts                              % This command is only needed if 
\usepackage{graphicx}
\usepackage{subfigure}
\usepackage{amsmath}
\usepackage{amsfonts,amssymb}
\usepackage{cite}
\usepackage{bm}
\usepackage{cases}
\usepackage[utf8]{inputenc}
\usepackage{makecell}
\usepackage{hyperref}

\usepackage{soul}
\usepackage{color, xcolor} 

\usepackage{algorithm}
\usepackage{algpseudocode}
\usepackage{amsmath}
\usepackage{graphics}
\usepackage{epsfig}
\usepackage{pifont}

\title{\LARGE \bf
Simple But Effective: Rethinking the Ability of Deep Learning in fNIRS to Exclude Abnormal Input
}

\author{Zhihao Cao$^{\dag}$% <-this % stops a space
\thanks{Zhihao Cao is with the Department of Mathematics, 
        ETH Zurich, Switzerland. Email: {\tt\small zhicao@ethz.ch}}% 
\thanks{$^{\dag}$Corresponding author. Email: {\tt\small zhicao@ethz.ch}}% 
\thanks{$^{\ddag}$\href{https://github.com/WhiteFireFox/fNIRS-MetricNet}{https://github.com/WhiteFireFox/fNIRS-MetricNet}}%
}

\begin{document}

\maketitle
\thispagestyle{empty}
\pagestyle{empty}

%%%%%%%%%%%%%%%%%%%%%%%%%%%%%%%%%%%%%%%%%%%%%%%%%%%%%%%%%%%%%%%%%%%%%%%%%%%%%%%%
\begin{abstract}

Functional near-infrared spectroscopy (fNIRS) is a non-invasive technique for monitoring brain activity. To better understand the brain, researchers often use deep learning to address the classification challenges of fNIRS data. Our study shows that while current networks in fNIRS are highly accurate for predictions within their training distribution, they falter at identifying and excluding abnormal data which is out-of-distribution, affecting their reliability. We propose integrating metric learning and supervised methods into fNIRS research to improve networks capability in identifying and excluding out-of-distribution outliers. This method is simple yet effective. In our experiments, it significantly enhances the performance of various networks in fNIRS, particularly transformer-based one, which shows the great improvement in reliability. We will make our experiment data available on GitHub$^{\ddag}$.

\end{abstract}

%%%%%%%%%%%%%%%%%%%%%%%%%%%%%%%%%%%%%%%%%%%%%%%%%%%%%%%%%%%%%%%%%%%%%%%%%%%%%%%%
\section{INTRODUCTION}

Functional Near-Infrared Spectroscopy (fNIRS) is valued for its non-invasive method of observing brain activity, emphasizing benefits such as portability and reduced sensitivity to electrical noise and motion artifacts \cite{naseer2015fnirs}. It measures shifts in oxygenated (HbO) and deoxygenated hemoglobin (HbR) levels through near-infrared light absorption, facilitating human brain function studies \cite{jobsis1977noninvasive}. Understanding fNIRS signals, particularly for classifying human behavioral intentions associated with these signals, shows promise for brain-computer interface (BCI) development \cite{eastmond2022deep}, potentially offering substantial support to individuals with disabilities \cite{ferrari2012brief}. This approach is one of various methods for interpreting brain signals, alongside electroencephalogram (EEG) \cite{buzsaki2012origin} and functional magnetic resonance imaging (fMRI) \cite{logothetis2008we}.

In fNIRS signal classification, two main approaches are traditional machine learning and deep learning. Traditional methods, such as Linear Discriminant Analysis (LDA) by Shin et al. \cite{shin2016open} and the feature extraction technique based on the general linear model (GLM) by Chen et al. \cite{chen2020classification}, often require manual feature selection and a deep understanding of the domain, which may limit their applicability. To address these challenges, recent progress has increasingly exploited deep learning's capabilities. Lyu et al. \cite{lyu2021domain} utilized CNN and LSTM techniques for fNIRS signal feature extraction, marking a move towards automated feature extraction. Similarly, Sun et al. adopted a 1D-CNN framework \cite{sun2020novel} to enhance feature extraction and classification accuracy of NIRS signals. Rojas et al.'s use of a Bi-LSTM model \cite{rojas2021pain} for automatic feature extraction from raw fNIRS data further demonstrates this evolution. Wang et al. \cite{wang2022transformer} introduced a transformer-based classification network (fNIRS-T), improving data utilization and network representation by analyzing spatial and channel-level signal features. This method signifies a significant advancement in classification efficacy. Furthermore, Wang et al.'s fNIRSNet \cite{wang2023rethinking}, which is well-calibrated \cite{cao2024calibration}, integrating delayed hemodynamic response into fNIRS classification, exemplifies a straightforward yet effective model, further boosting classification precision.

\begin{figure}[htpb]
  \vspace{-0.3cm}
  \centering
  \vspace{0.1cm}
  \includegraphics[width=3.3in]{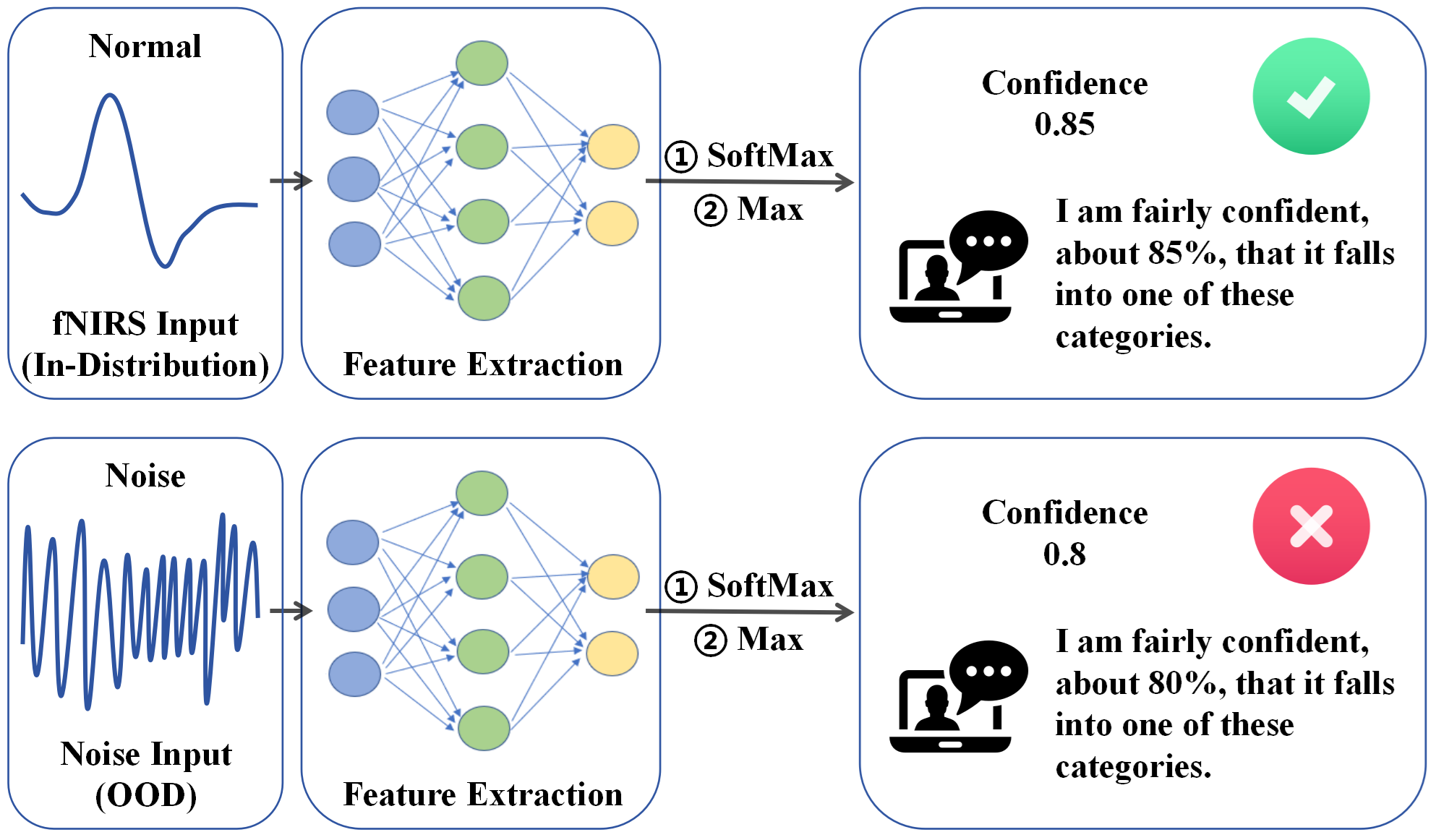}
  \vspace{-0.1cm}
  \caption{In each cross-validation, inputing the in-distribution data or out-of-distribution data (e.g. meaningless noise), and observing the confidence of the output after the SoftMax. Surprisingly, obtaining high confidence under normal data and noise input conditions, which shows the network's inability to exclude meaningless noise input.}
  \label{fig:intro_ood}
  \vspace{-0.2cm}
\end{figure}

While today's deep learning networks for fNIRS exhibit high predictive accuracy, we are surprised to find these models struggling with identifying and excluding meaningless noisy inputs. As depicted in the Figure\ref{fig:intro_ood}, models in fNIRS display high confidence for both normal inputs (in-distribution) and noisy inputs (out-of-distribution, OOD), indicating a lack of model credibility. Ideally, a deep learning model should not provide a high confidence answer for inputs from unknown classes, i.e., classes not included in its training. 

In other fields, to solve OOD problems, Hendrycks et al. \cite{hendrycks2016baseline} introduced a baseline for evaluating OOD detection methods, based on the assumption that OOD samples display greater confidence dispersion across classes than in-distribution samples. Liang et al. \cite{liang2017enhancing} developed ODIN, utilizing temperature scaling and input preprocessing to distinguish OOD samples. Moreover, In \cite{lee2017training}, enhancements to in-distribution sample accuracy were proposed, alongside mechanisms to reduce confidence in OOD samples by introducing two additional loss terms. These methods aim to refine the accuracy of confidence estimates in cross-entropy-based approaches, reducing their overconfidence in OOD instances. However, due to the nature of cross-entropy methods, which require the output vector representation to always sum to 1, these networks will still allocate the highest probability to the nearest class for inputs outside the training distribution \cite{masana2018metric}.

Therefore, Masana et al. \cite{masana2018metric} introduced a method based on metric learning to ensure that the inputs sharing the same label are close to each other in the embedding space, while the inputs with different labels maintain a minimum distance apart. These networks bypass the need for SoftMax layers, thus avoiding the compulsion to categorize the out-of-distributed inputs into known classes. Additionally, Mohseni et al. \cite{mohseni2020self} suggested training a distribution classifier and detector within a single network, eliminating the need for adjusting subsamples to match a targeted OOD set. This approach allows the use of any "free" unlabeled OOD dataset for training. Inspired by previous ideas in solving OOD problems in other fields, we find that integrating metric learning with supervised methods offers a straightforward yet effective solution for fNIRS to exclude OOD data, significantly enhancing the reliability of the system. The contributions of this letter are summarized as follows.

\begin{itemize}

\item In our experiments, we test and verify that most current deep learning models for fNIRS classification exhibit overconfidence when dealing with data outside the distribution, though they demonstrate high accuracy for in-distribution data.

\item We propose to develop a system based on metric learning and supervised methods to enhance the identification effect of out-of-distribution data. Our experiments will confirm its effectiveness across various models in fNIRS, particularly transformer-based one.

\end{itemize}

The rest of this letter is organized as follows. Section \ref{sec:dataset} introduces the dataset used in this research, including the preprocessing steps for the fNIRS dataset. Section \ref{sec:issues} discusses that various current deep learning models for fNIRS classification display overconfidence with OOD data, despite their high accuracy for in-distribution data. Section \ref{sec:network} outlines the key components of the network and details the training methodology. Section \ref{sec:result} presents the experiment's results and the model's performance within the framework of the proposed method, confirming its effectiveness in fNIRS classification. Section \ref{sec:conclusion} offers a summary and conclusion of the findings discussed in this letter.

\section{FUNCTIONAL NEAR-INFRARED SPECTROSCOPY DATASET \label{sec:dataset}}

To guarantee reproducibility and diversity in our research, we utilized two open-source datasets for our experiments. These datasets comprise a mental arithmetic classification task \cite{shin2016open} and a unilateral finger and foot tapping classification task \cite{bak2019open}. What' more, this section outlines the preprocessing methods applied to these datasets.

\subsection{MA Dataset and UFFT Dataset} 

\begin{figure}[htpb]
  \centering
  \vspace{-0.4cm}
  \subfigure[]{
  \includegraphics[width=1.55in]{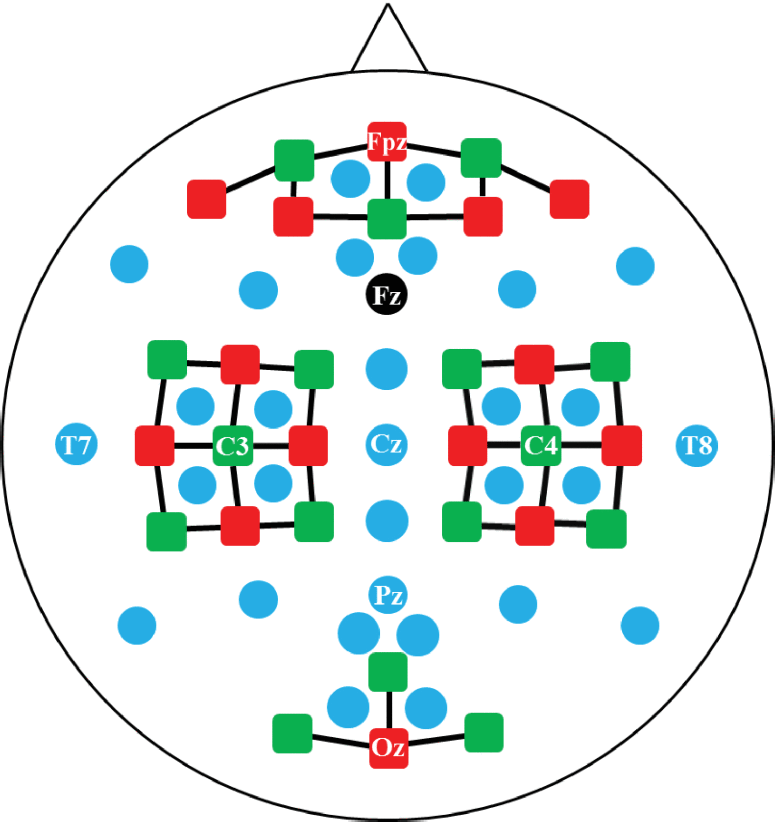}
  }
  \subfigure[]{
  \includegraphics[width=1.55in]{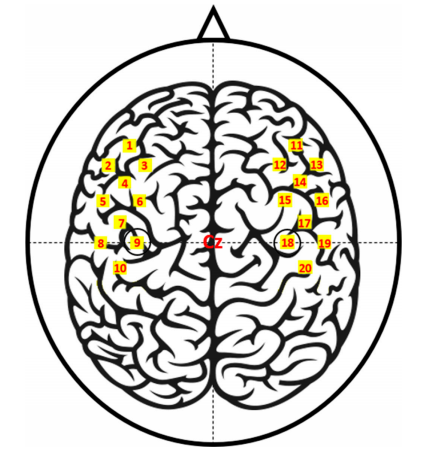}
  }
  \vspace{-0.1cm}
  \caption{(a) Sensor location layout for MA \cite{shin2016open}. (b) Sensor location layout for UFFT \cite{bak2019open}.}
  \label{fig:dataset}
  \vspace{-0.5cm}
\end{figure}

The Mental Arithmetic (MA) dataset was collected at a 12.5 Hz sampling rate \cite{shin2016open} with the NIRScout device. As shown in Figure \ref{fig:dataset} (a), it included fourteen sources and sixteen detectors. The structure of each trial was organized as follows: a 2-second introductory period, followed by a 10-second task period, and a 15 to 17-second inter-trial rest, ending with a brief 250 ms beep. Subjects participated in tasks displayed randomly on a screen, consisting of mental arithmetic and baseline activities.

The Unilateral Finger and Foot Tap (UFFT) dataset was captured using a three-wavelength, continuous-time, multi-channel fNIRS system equipped with transmitters (Tx) and receivers (Rx), as illustrated in Figure \ref{fig:dataset} (b) \cite{bak2019open}. The protocol for each trial included a 2-second introduction, a 10-second task period, and a rest period ranging from 17 to 19 seconds between trials. Participants were directed to execute three specific actions—Right Hand Tap (RHT), Left Hand Tap (LHT), and Foot Tap (FT)—as prompted by instructions displayed on a screen during the task segment.

\subsection{Signal Preprocessing} 

In the preprocessing phase of the Mental Arithmetic (MA) dataset, changes in the concentrations of deoxygenated and oxygenated hemoglobin (HbR and HbO, respectively) are computed using the modified Beer-Lambert law \cite{cope1988system}. Then, HbO and HbR data are filtered by a sixth-order zero-phase Butterworth band-pass filter with a filtering range of 0.01 to 0.1 Hz. The data are then segmented with a sliding window approach (3-second window size and 1-second step size) \cite{shin2016open}, and underwent normalization.

In the Unilateral Finger and Foot Tap (UFFT) dataset, the filtering of HbO and HbR data is achieved through a third-order zero-phase Butterworth band-pass filter, operating within the 0.01 to 0.1 Hz range. Following this, baseline correction is applied by deducting the mean value of the data from the -1 to 0 second period before the task initiation. The dataset is further processed by segmenting the data using a 3-second sliding window with a 1-second step \cite{sun2020novel}, culminating in data normalization.

\section{EXISTING PROBLEMS  \label{sec:issues}}

In this section, we aim to experimentally demonstrate that the current networks in fNIRS struggles to accurately exclude OOD data (e.g., meaningless noise). To do this, following the approach suggested by Wang et al. \cite{wang2023rethinking}, we adopt wider and clearer criteria that encompass subject-specific measures. Subject-specific analysis involves training a deep learning model for each participant using 5-fold cross-validation. This method divides the dataset into training and validation sets based on trials to prevent information leakage.

To confirm that many existing deep learning models for fNIRS classification tasks still display overconfidence with OOD data, we conduct a thorough evaluation with 5-fold cross-validation. This process involves assessing the model's accuracy and confidence on the in-distribution validation sets, as well as its confidence when encountering OOD data in fNIRS, such as generated meaningless noise. The ultimate experimental result is the average performance across all subject test sets, as shown in Table \ref{tab:ma_confidence} and \ref{tab:ufft_confidence}.

\begin{table}[hptb]
\vspace{-0.1cm}
\caption{VALIDATION ON MA DATASET}
\label{tab:ma_confidence}
\vspace{-0.3cm}
\begin{center}
\begin{tabular}{c|c|c|c}
\hline
  & Accuracy $\uparrow$ & \makecell{Confidence $\uparrow$ \\ @ In-Distribution} & \textcolor{red}{\makecell{Confidence $\approx \frac{1}{n_{class}}$ \\ @ Out-of-Distribution}} \\
\hline
CNN  & 0.61 ± 0.19 & 0.77 ± 0.05 & 0.65 ± 0.16 (little bad) \\
\hline
1D-CNN  & 0.64 ± 0.13 & 0.89 ± 0.02 & 0.82 ± 0.09 (bad) \ding{55}  \\
\hline
LSTM  & 0.57 ± 0.12 & 0.76 ± 0.07 & 0.78 ± 0.12 (bad) \ding{55} \\
\hline
fNIRS-T  & 0.65 ± 0.14 & 0.89 ± 0.04 & 0.81 ± 0.12 (bad) \ding{55} \\
\hline
fNIRSNet  & 0.72 ± 0.14 & 0.82 ± 0.04 & 0.89 ± 0.03 (bad) \ding{55} \\
\hline
\end{tabular}
\end{center}
\vspace{-0.4cm}
\end{table}

\begin{table}[hptb]
\vspace{-0.1cm}
\caption{VALIDATION ON UFFT DATASET}
\label{tab:ufft_confidence}
\vspace{-0.3cm}
\begin{center}
\begin{tabular}{c|c|c|c}
\hline
  & Accuracy $\uparrow$ & \makecell{Confidence $\uparrow$ \\ @ In-Distribution} & \textcolor{red}{\makecell{Confidence $\approx \frac{1}{n_{class}}$ \\ @ Out-of-Distribution}} \\
\hline
CNN  & 0.65 ± 0.14 & 0.82 ± 0.04 & 0.67 ± 0.10 (bad) \ding{55} \\
\hline
1D-CNN  & 0.57 ± 0.17 & 0.86 ± 0.03 &  0.79 ± 0.11 (bad) \ding{55} \\
\hline
LSTM  & 0.46 ± 0.14 & 0.79 ± 0.05 & 0.78 ± 0.14 (bad) \ding{55} \\
\hline
fNIRS-T & 0.62 ± 0.18 & 0.81 ± 0.05 & 0.71 ± 0.13 (bad) \ding{55} \\
\hline
fNIRSNet & 0.69 ± 0.20 & 0.67 ± 0.06 & 0.73 ± 0.05 (bad) \ding{55} \\
\hline
\end{tabular}
\end{center}
\vspace{-0.4cm}
\end{table}

The results show the model retains high confidence in out-of-distribution data, which is ill-conditioned. Ideally, the model with SoftMax-Layer should distribute confidence evenly across known distributions, indicating uncertainty about the real distribution of the out-of-distribution data.

\section{NETWORK STRUCTURE AND TRAINING \label{sec:network}}

In this section, we detail the network architecture and training methods we devised. Our objective is to enable the same classifier, trained on in-distribution data, to exclude outliers from OOD. To achieve this, our approach integrates ``OOD generator'', ``detector'' and ``subspace'' directly into the deep learning model in fNIRS, specifically trained to identify and exclude OOD data.

\begin{figure}[htpb]
  \centering
  \includegraphics[width=3.3in]{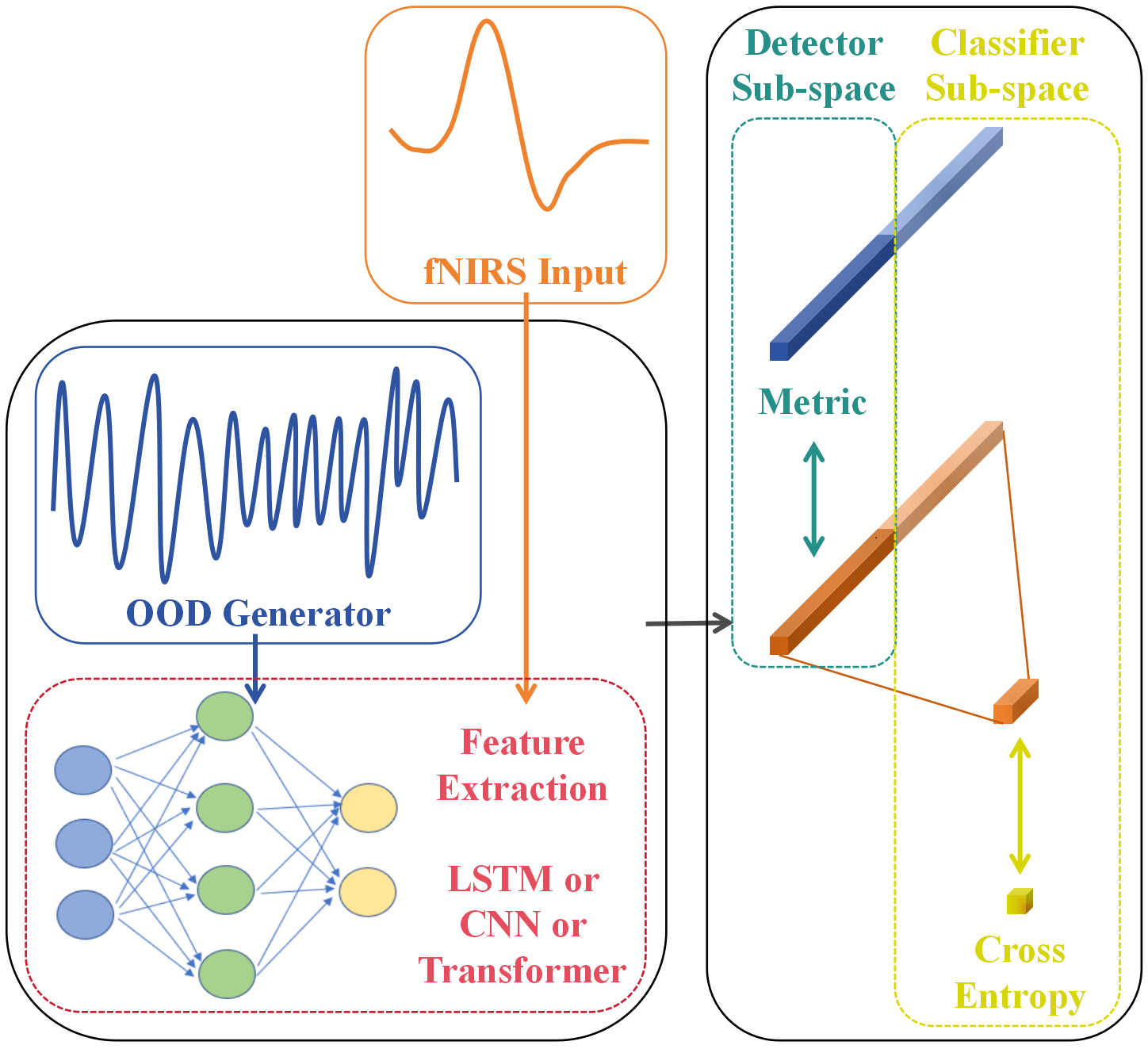}
  \vspace{-0.1cm}
  \caption{The structure of neural network.}
  \label{fig:network}
  \vspace{-0.5cm}
\end{figure}

Despite its conceptual simplicity, our method demonstrates significant effectiveness in excluding OOD data through comprehensive experiments, particularly with transformer-based models, showing its performance.

\subsection{OOD Generator} 

The OOD generator primarily creates unlabeled OOD data as negative examples. This generator operates independently, often utilizing real-world OOD data. For instance, in MA missions, OOD data could be \textbf{unlabeled} fNIRS data from different mission contexts with the same instruments. Due to the absence of such conditions in our studies, we produce Gaussian white noise, a common real-world phenomenon, as OOD data. This approach is practical because, in deploying fNIRS, the presence of significant Gaussian white noise is expected. Consequently, it's crucial for the neural network model in fNIRS to identify and discard these OOD noises.

\subsection{Feature Extraction} 

Feature extraction transforms raw data into meaningful representations by abstracting essential features, facilitating the exploration of data's hierarchical structure \cite{lamer2022standardized}. This process is particularly beneficial for identifying spatiotemporal characteristics in fNIRS data, improving model comprehension, and enhancing generalization abilities. Previous studies have demonstrated the effective feature extraction capabilities of backbone networks in CNN \cite{sun2020novel}, LSTM \cite{lyu2021domain}, transformer \cite{wang2022transformer}, and fNIRSNet \cite{wang2023rethinking}. Our experiments reveal that models incorporating transformer architectures significantly enhance performance in identifying and excluding OOD data.

\subsection{Detector Subspace}

We adopt the concept of subspace, partitioning the feature space into two distinct areas: detection subspace and classification subspace. Initially focusing on the detection subspace, it is constituted by the initial 64 dimensions of the feature vector, serving to identify and eliminate OOD data. Unlike traditional softmax calibration techniques \cite{lee2017training, hendrycks2016baseline}, by evaluating within this subspace, we can widen the gap between the feature vectors of in-distribution data and those of OOD data in detection feature space, facilitating feature differentiation. The formulation of metric loss within the detection subspace is expressed as

\begin{equation}
L_{metric} = \mathbb{E}[-metric(net(\hat{x}_{in})[:k], net(\hat{x}_{out})[:k])]\;,
\end{equation}
where $\hat{x}_{in}$ and $\hat{x}_{out}$ are in-distribution data and OOD data, respectively, $net(\cdot)$ is to map the data to a feature vector and $net(\cdot)[:k]$ refers to extracting the first $k$ dimensions of the feature vector, with $k=63$. Moreover, $metric(\cdot)$ represents a generalized distance function, with the cosine distance function being our choice for measurement. It's important to note that a smaller cosine value indicates a greater distance within the feature space, a relationship that inversely applies if using the Euclidean distance function. Hence, our description pertains to distance in a broader sense, necessitating the inclusion of a negative sign to accommodate this interpretation.

\begin{figure}[htpb]
  \centering
  \vspace{-0.5cm}
  \subfigure[Cosine Distance]{
  \includegraphics[width=1.55in]{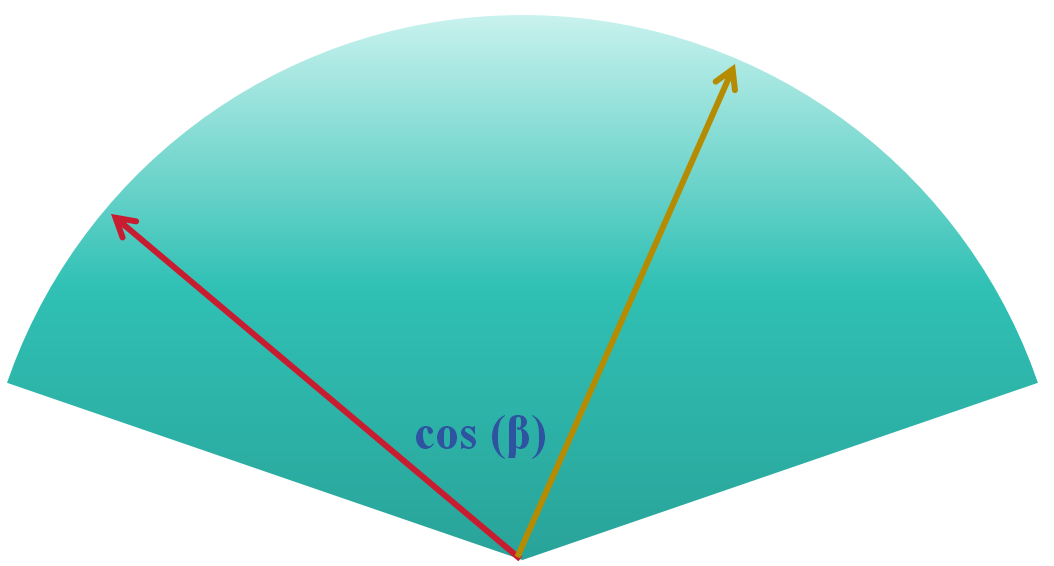}
  }
  \subfigure[Euclidean Distance]{
  \includegraphics[width=1in]{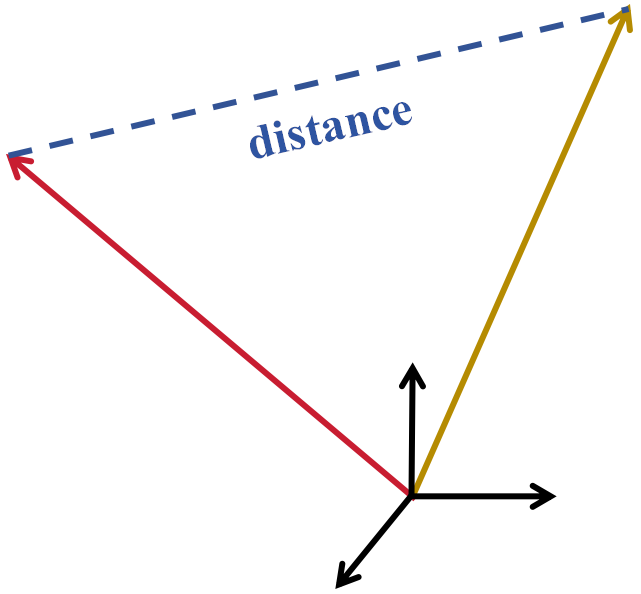}
  }
  \vspace{-0.1cm}
  \caption{(a) Explicit expression of cosine distance function. (b) Explicit expression of Euclidean distance function.}
  \label{fig:distance}
  \vspace{-0.5cm}
\end{figure}

\subsection{Classifier Subspace}

The classification subspace is designed to differentiate among the categories of in-distribution data, leveraging the last 64 dimensions of the feature vector for classification learning. Through a nonlinear transformation of this 64-dimensional feature vector, we obtain a feature vector corresponding to the number of categories. Subsequently, cross-entropy loss is computed to fulfill the objective of classifying in-distribution data effectively. The classification loss function is calculated as

\begin{equation}
L_{class} = \mathbb{E}_{P_{in(\hat{x}_{in},\hat{y}_{in})}}[-log(P_{\theta}(y=\hat{y}_{in} | \hat{x}_{in})]\;,
\end{equation}
where $\hat{x}_{in}$ represents the input data of in-distribution samples, while $\hat{y}_{in}$ denotes their corresponding labels.

\subsection{Two-stage Training}

We adopt a two-step training approach that begins with fully supervised learning of distribution features, followed by the incorporation of metric learning to improve the network's capability to exclude OOD data. The two-step training methodology is detailed in Algorithm \ref{alg:two_stage}. In the subsequent sections, we will discuss the routines involved in this algorithm and provide guidance on effective training practices.

\begin{algorithm}
\caption{Two-stage Training Process \label{alg:two_stage}}
\textbf{Procedure} Supervised Learning \newline
\mbox{\quad} \textbf{Input} Batch of In-distribution data from different classes \newline
\mbox{\quad} \textbf{Loss} $L = \mathbb{E}_{P_{in(\hat{x}_{in},\hat{y}_{in})}}[-log(P_{\theta}(y=\hat{y}_{in} | \hat{x}_{in})]$ \newline

\textbf{Procedure} Metric Learning \newline
\mbox{\quad} \textbf{Input} Batch of In-distribution data from different classes \newline
\mbox{\quad} \textbf{Generate} Batch of OOD data. \newline
\mbox{\quad} \textbf{Loss$_1$} $L = \mathbb{E}_{P_{in(\hat{x}_{in},\hat{y}_{in})}}[-log(P_{\theta}(y=\hat{y}_{in} | \hat{x}_{in})]$ \newline
\mbox{\quad} \textbf{Loss$_2$} $L = \mathbb{E}[-metric(net(\hat{x}_{in})[:k], net(\hat{x}_{out})[:k])]$
\end{algorithm}

The first stage of training emphasizes fully supervised learning of in-distribution features. This phase is flexible, allowing for any suitable network model in fNIRS and training duration to achieve the desired classification accuracy. The focus here is on utilizing in-distribution training data to minimize label loss, for which we employ cross-entropy loss exclusively. During the second training stage, our OOD generator begins producing a substantial volume of unlabeled OOD data. These OOD samples are fed into the network to generate feature vector $net(\hat{x}_{in})$, while in-distribution data are processed to produce feature vector $net(\hat{x}_{out})$. The objective is for deep learning algorithms to concurrently optimize both cross-entropy loss and metric loss. This dual optimization ensures effective rejection of OOD data without compromising the model's accuracy. The optimization goal is articulated as

\begin{equation}
L = L_{class} + \lambda L_{metric}\;,
\end{equation}
where $\lambda$ is a hyperparameter that adjusts based on the network's feature extraction capability.

\section{RESULTS \label{sec:result}}

\subsection{Visualization of Feature Vectors in the Detector Subspace}

A practical method to assess the efficacy of an OOD detector involves visualizing the feature vectors produced by the deep learning network within the detector's feature subspace.

\begin{figure}[htpb]
  \centering
  \subfigure[\textbf{Evaluation} on MA Dataset]{
  \includegraphics[width=1.55in]{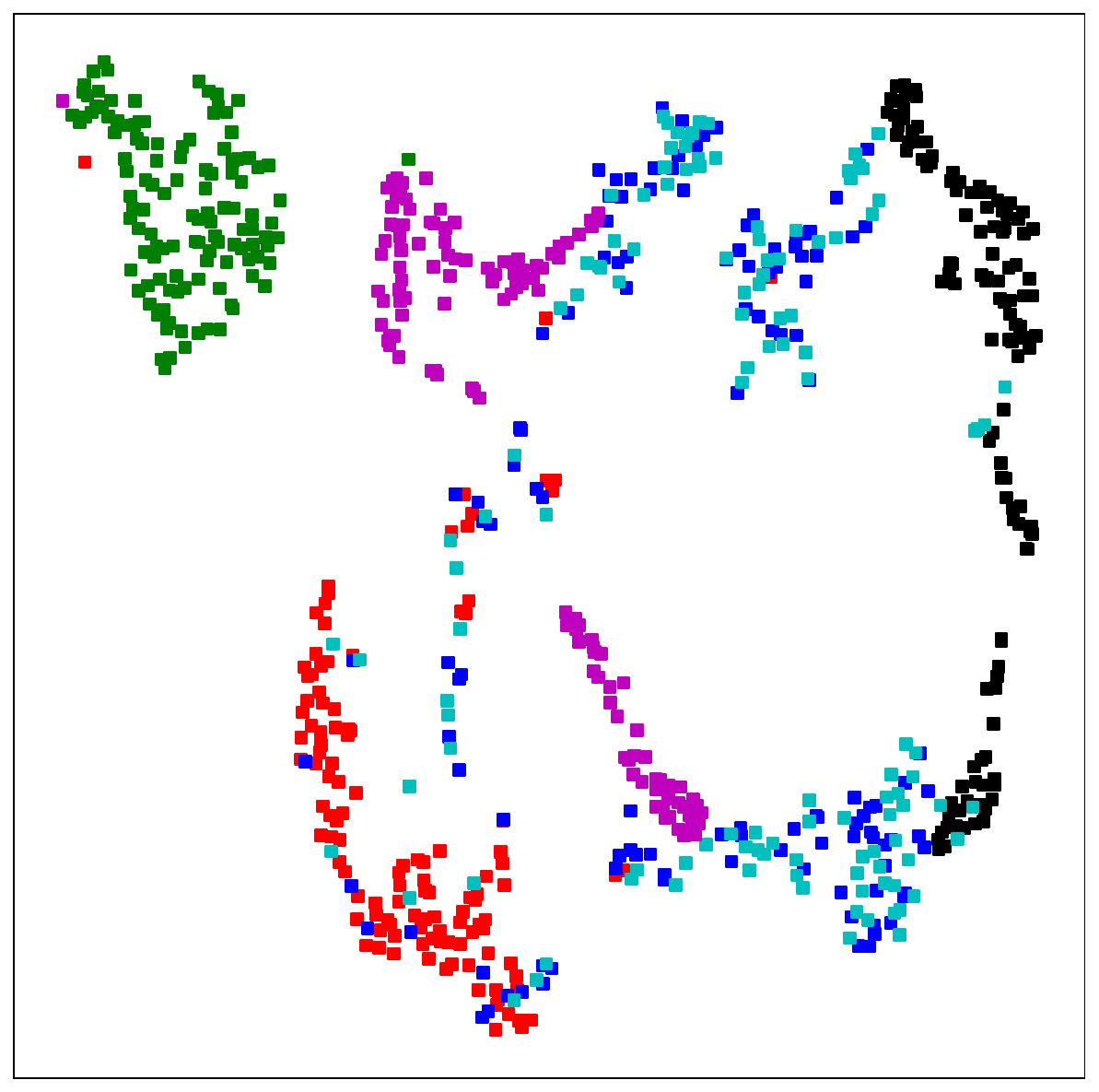}
  }
  \subfigure[\textbf{Evaluation} on UFFT Dataset]{
  \includegraphics[width=1.55in]{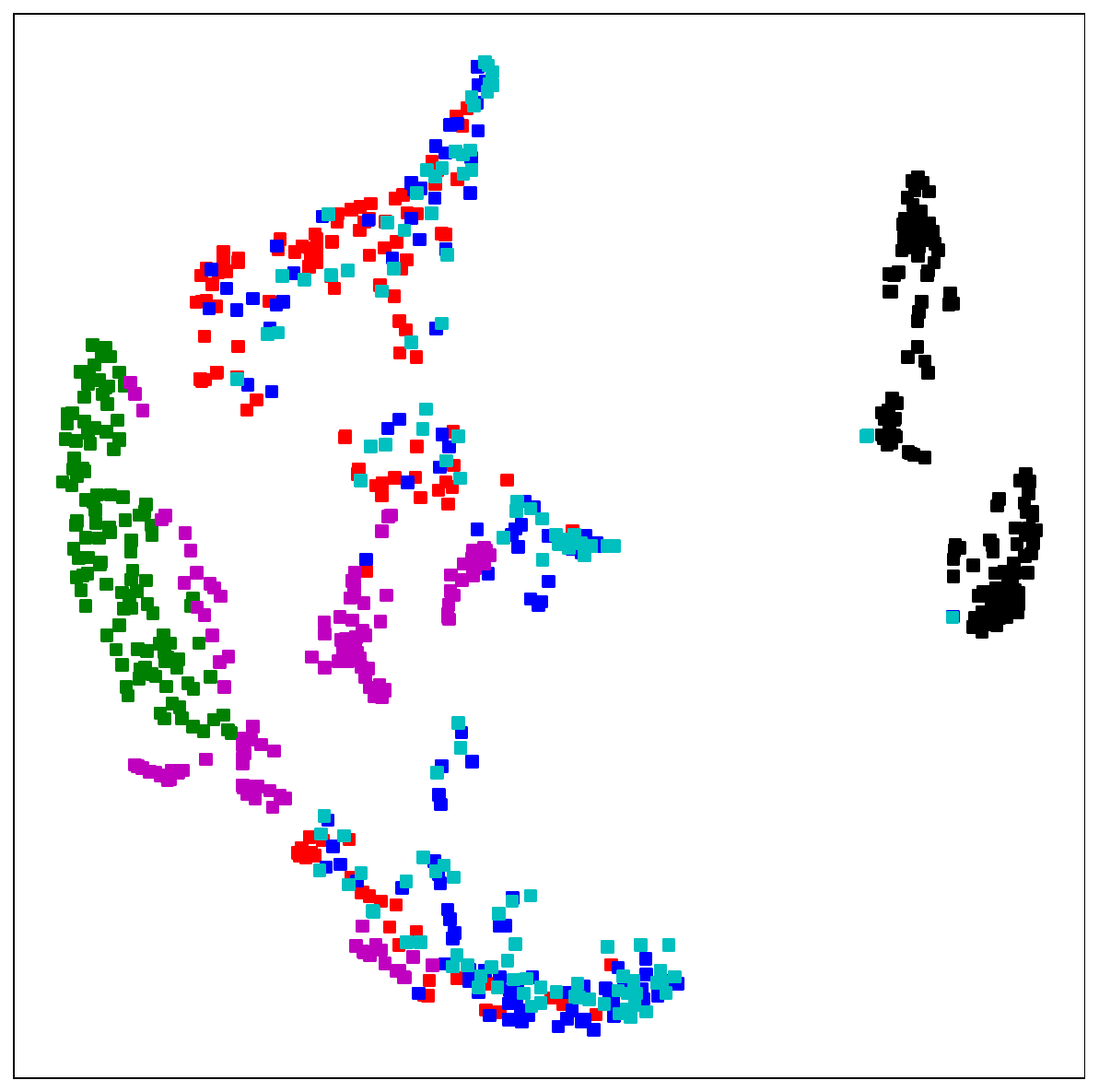}
  }
  \vspace{-0.1cm}
  \caption{Visualization of the results of fNIRS-Transformer as feature extraction on the detector feature subspace.}
  \label{fig:fnirst}
  \vspace{-0.5cm}
\end{figure}

\begin{figure}[htpb]
  \centering
  \subfigure[\textbf{Evaluation} on MA Dataset]{
  \includegraphics[width=1.55in]{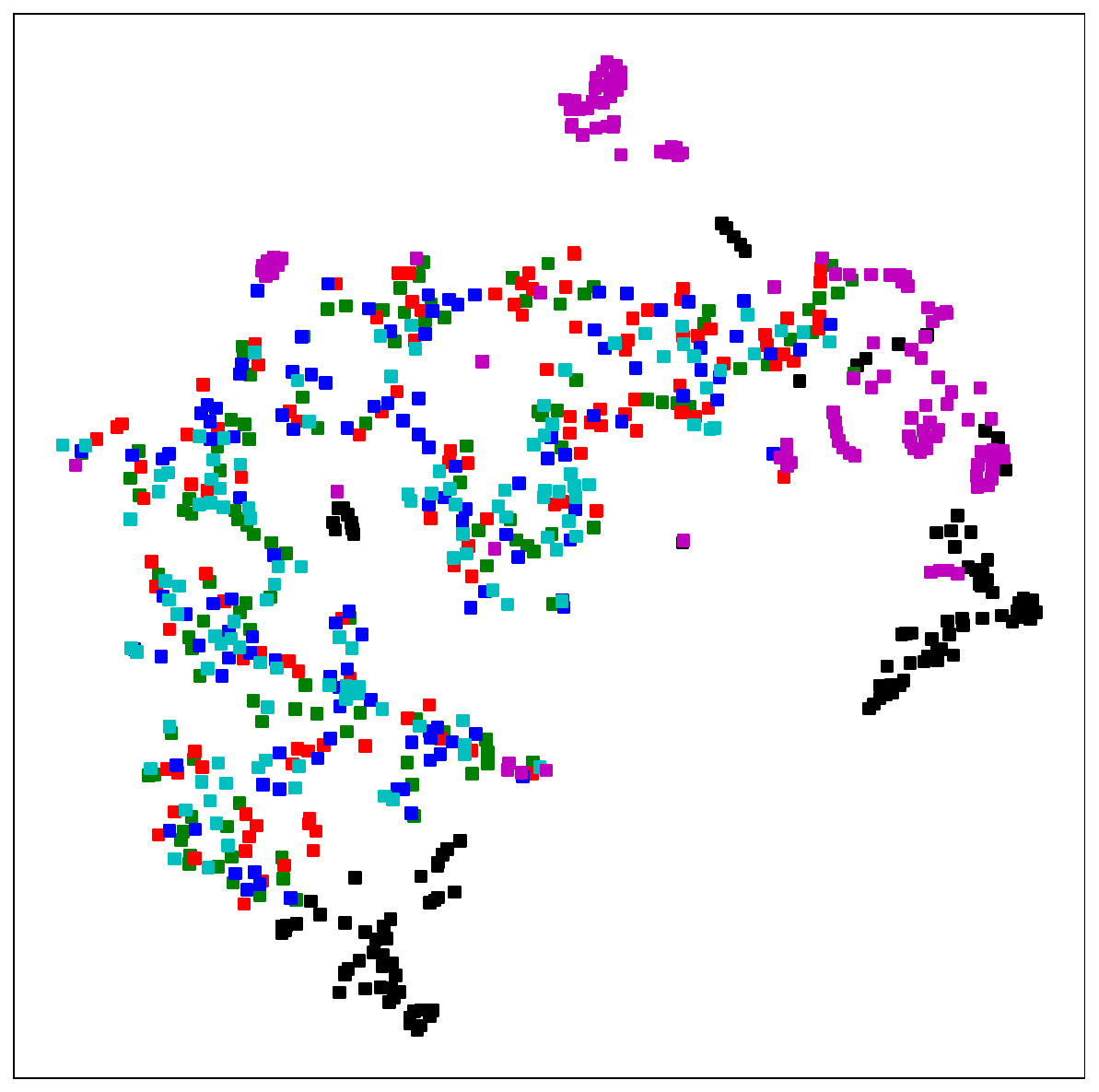}
  }
  \subfigure[\textbf{Evaluation} on UFFT Dataset]{
  \includegraphics[width=1.55in]{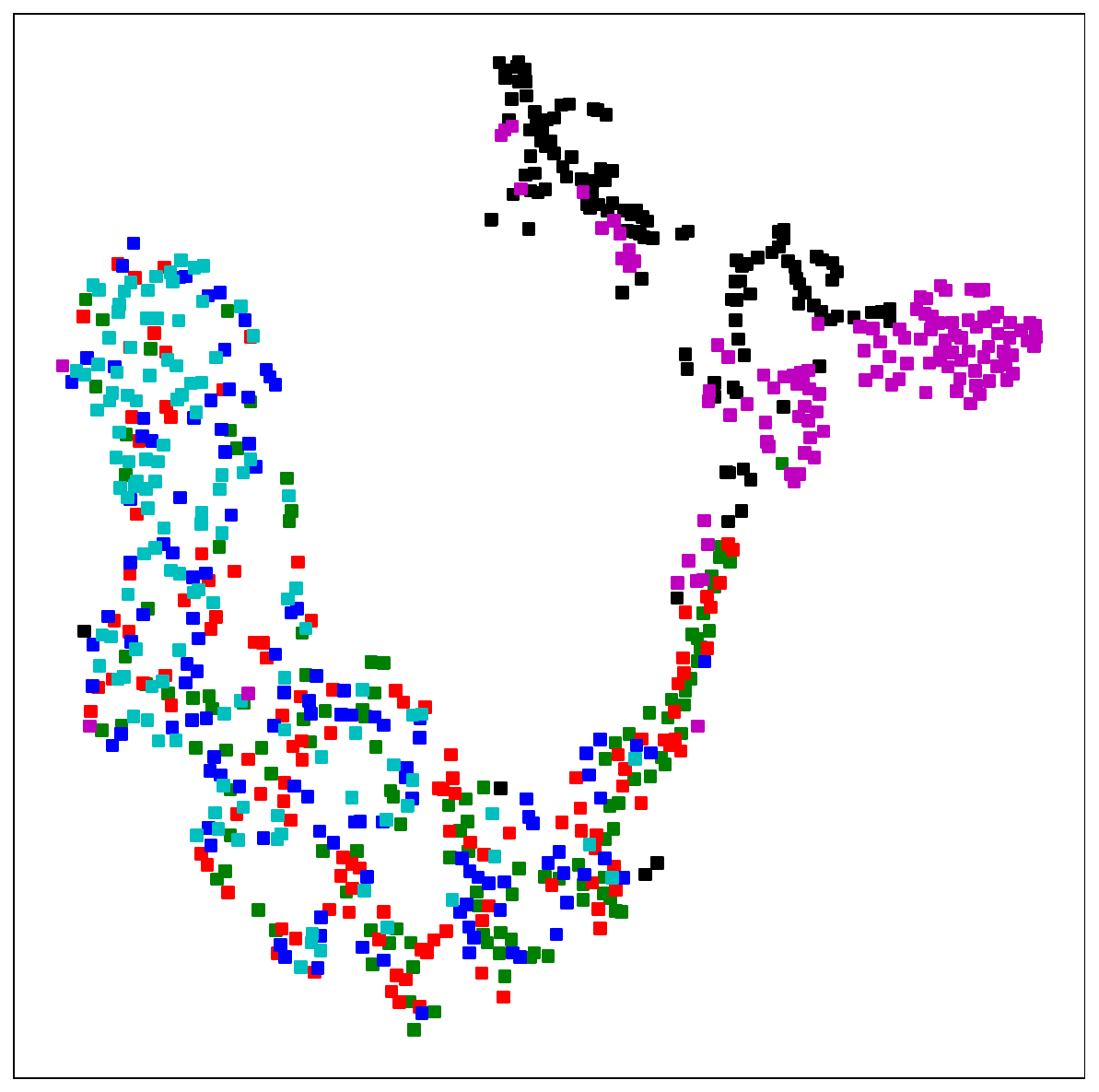}
  }
  \vspace{-0.1cm}
  \caption{Visualization of the results of fNIRSNet as feature extraction on the detector feature subspace.}
  \label{fig:fnirsnet}
  \vspace{-0.5cm}
\end{figure}

\begin{figure}[htpb]
  \centering
  \subfigure[\textbf{Evaluation} on MA Dataset]{
  \includegraphics[width=1.55in]{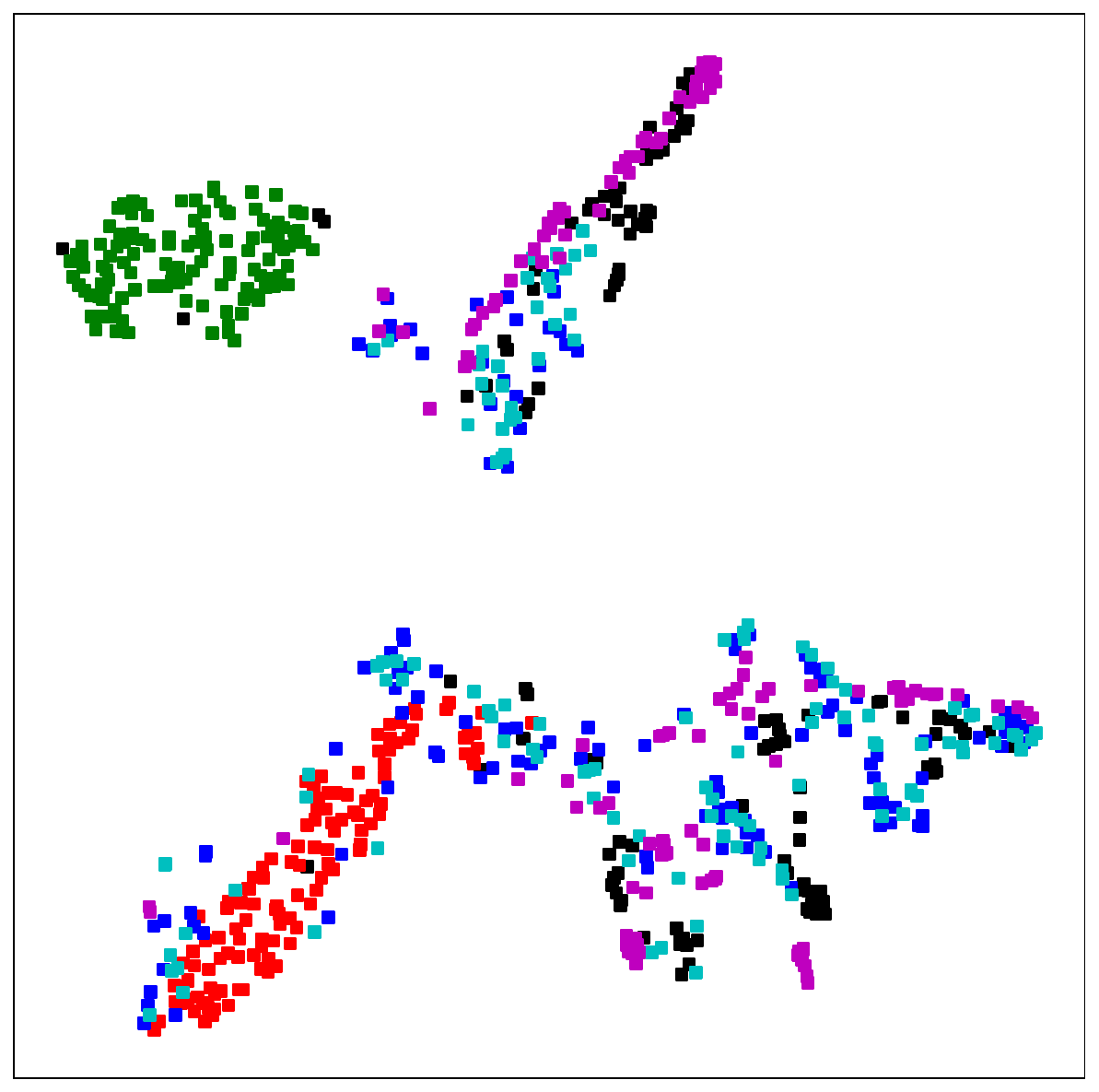}
  }
  \subfigure[\textbf{Evaluation} on UFFT Dataset]{
  \includegraphics[width=1.55in]{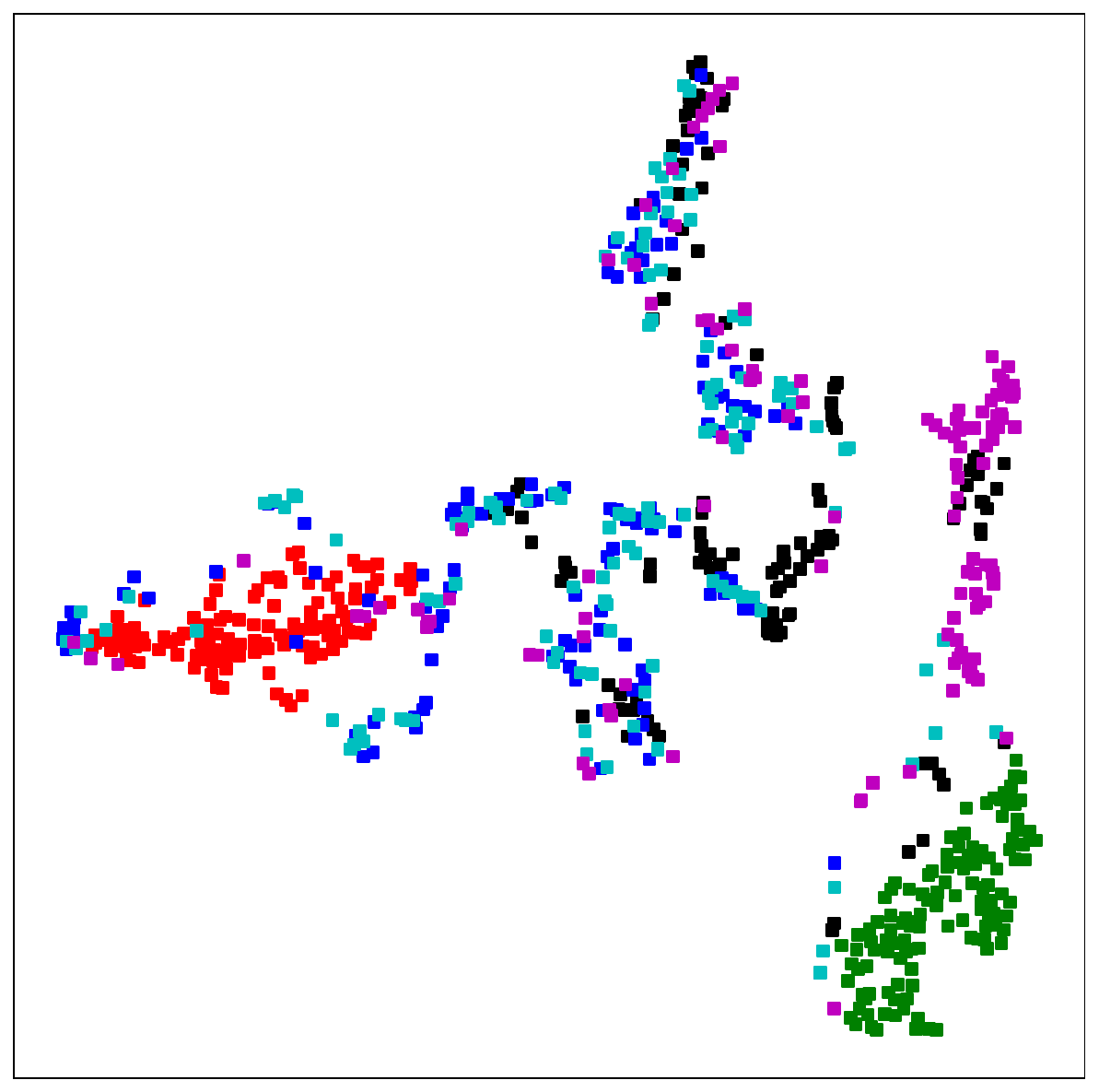}
  }
  \vspace{-0.1cm}
  \caption{Visualization of the results of CNN as feature extraction on the detector feature subspace. In Figures \ref{fig:fnirst}, \ref{fig:fnirsnet}, and \ref{fig:cnn_result}, black denotes in-distribution data, green represents random white noise, red signifies Gaussian white noise, blue indicates in-distribution data with substantial noise (requiring exclusion), cyan denotes in-distribution data with a moderate level of noise (most of which should be excluded), and purple illustrates the reorganization of data channels from in-distribution data to simulate potential real-world out-of-distribution scenarios (high difficulty).}
  \label{fig:cnn_result}
  \vspace{-0.5cm}
\end{figure}

\begin{table*}[hptb]
\vspace{0.2cm}
\caption{RESULT SCORE OF MA DATASET}
\label{tab:ma_result}
\vspace{-0.3cm}
\begin{center}
\begin{tabular}{c|c|c|c|c|c}
\hline
  & Accuracy $\uparrow$ & \makecell{Confidence $\uparrow$ \\ @ In-Distribution} & \makecell{Acceptance Rate $\uparrow$ \\ @ In-Distribution} & \makecell{Exclusion Rate $\uparrow$ \\ @ OOD (Noise, Low Difficulty)}  & \makecell{Exclusion Rate $\uparrow$ \\ @ OOD (High Difficulty)} \\
\hline
CNN  & 0.61 ± 0.19 & 0.77 ± 0.05 & \ding{55} & \ding{55} & \ding{55}  \\
\hline
\textbf{CNN-Metric}  & 0.70 ± 0.14 & 0.94 ± 0.01 & 0.88 ± 0.09 & 0.97 ± 0.14 & 0.13 ± 0.12  \\
\hline
fNIRS-T  & 0.65 ± 0.14 & 0.89 ± 0.04 & \ding{55} & \ding{55} & \ding{55}  \\
\hline
\textbf{fNIRS-T-Metric} &  0.65 ± 0.14 & 0.92 ± 0.02 & \textbf{1.00 ± 0.00} & \textbf{1.00 ± 0.00} & \textbf{0.47 ± 0.21} \\
\hline
fNIRSNet  & 0.72 ± 0.14 & 0.82 ± 0.04  & \ding{55} & \ding{55} & \ding{55}  \\
\hline
\textbf{fNIRSNet-Metric}  & \textbf{0.73 ± 0.11} & \textbf{0.91 ± 0.02} & 0.51 ± 0.17 & 0.82 ± 0.16 & 0.45 ± 0.13  \\
\hline
\end{tabular}
\end{center}
\vspace{-0.2cm}
\end{table*}

\begin{table*}[hptb]
\vspace{-0.1cm}
\caption{RESULT SCORE OF UFFT DATASET}
\label{tab:ufft_result}
\vspace{-0.3cm}
\begin{center}
\begin{tabular}{c|c|c|c|c|c}
\hline
  & Accuracy $\uparrow$ & \makecell{Confidence $\uparrow$ \\ @ In-Distribution} & \makecell{Acceptance Rate $\uparrow$ \\ @ In-Distribution} & \makecell{Exclusion Rate $\uparrow$ \\ @ OOD (Noise, Low Difficulty)}  & \makecell{Exclusion Rate $\uparrow$ \\ @ OOD (High Difficulty)} \\
\hline
CNN  & 0.65 ± 0.14 & 0.82 ± 0.04 & \ding{55} & \ding{55} & \ding{55}  \\
\hline
\textbf{CNN-Metric}  & 0.65 ± 0.14 & 0.91 ± 0.03 & 0.77 ± 0.10 & 0.87 ± 0.22 & 0.26 ± 0.10  \\
\hline
fNIRS-T  & 0.62 ± 0.18 & 0.81 ± 0.05 & \ding{55} & \ding{55} & \ding{55}  \\
\hline
\textbf{fNIRS-T-Metric} & 0.58 ± 0.19 & \textbf{0.87 ± 0.03} & \textbf{1.00 ± 0.00} & \textbf{1.00 ± 0.00} & \textbf{0.68 ± 0.22} \\
\hline
fNIRSNet  & \textbf{0.69 ± 0.20} & 0.67 ± 0.06 & \ding{55} & \ding{55} & \ding{55}  \\
\hline
\textbf{fNIRSNet-Metric}  & 0.65 ± 0.15 & 0.85 ± 0.03 & 0.48 ± 0.16 & 0.74 ± 0.21 & 0.47 ± 0.12 \\
\hline
\end{tabular}
\end{center}
\vspace{-0.6cm}
\end{table*}

The detector's effectiveness is directly linked to the spatial separation between in-distribution and OOD feature vectors within the feature subspace. A larger distance signifies enhanced detection and rejection of OOD data. We conduct detailed experiments across various networks, with results documented in Table \ref{tab:ma_result} and \ref{tab:ufft_result}. These experiments reveal fascinating insights, depicted in Figure \ref{fig:fnirst}, \ref{fig:fnirsnet}, and \ref{fig:cnn_result}. 

Our results indicate that the model, trained through our approach, is proficient in nearly completely excluding OOD data as well as instances with a significant mix of noise. This outcome is highly significant. In practical scenarios, if a deep learning model can identify and report the presence of substantial noise, it would markedly enhance the model's reliability. Furthermore, our analysis reveal that networks based on the transformer architecture yield the most effective results, essentially excluding the majority of OOD data. Conversely, we hypothesize that the fNIRSNet network exhibits a weak capability in metric feature extraction, evidenced by its struggle to differentiate between ostensibly plausible reorganized data and genuine in-distribution outputs.

\subsection{Visualization of Feature Vectors in the Classifier Subspace}

\begin{figure}[htpb]
  \centering
  \vspace{-0.2cm}
  \subfigure[\textbf{Evaluation} on MA Dataset \newline @ \qquad Detector Subspace]{
  \includegraphics[width=1.55in]{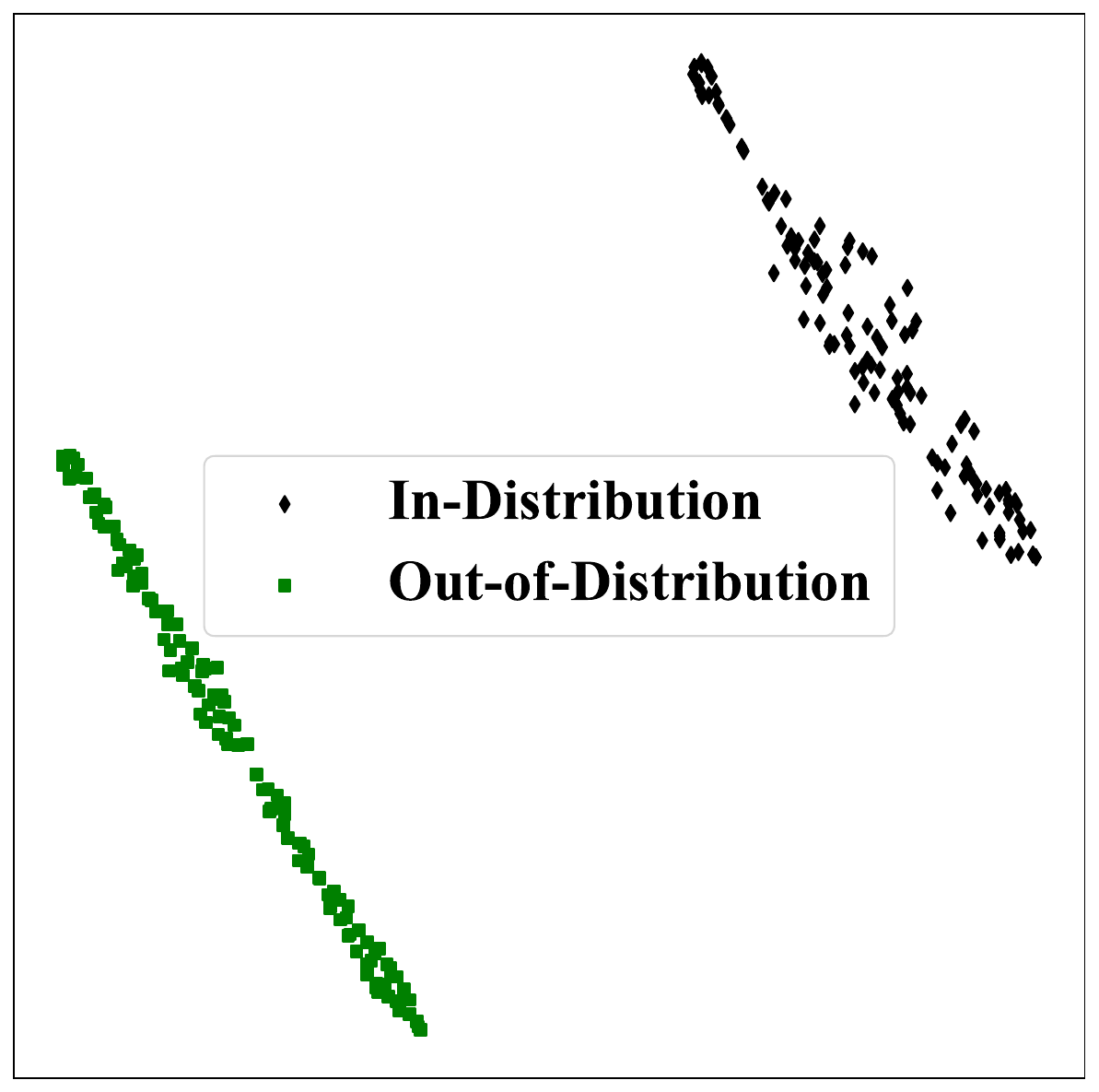}
  }
  \subfigure[\textbf{Evaluation} on UFFT Dataset  \newline @ \quad \quad Detector Subspace]{
  \includegraphics[width=1.55in]{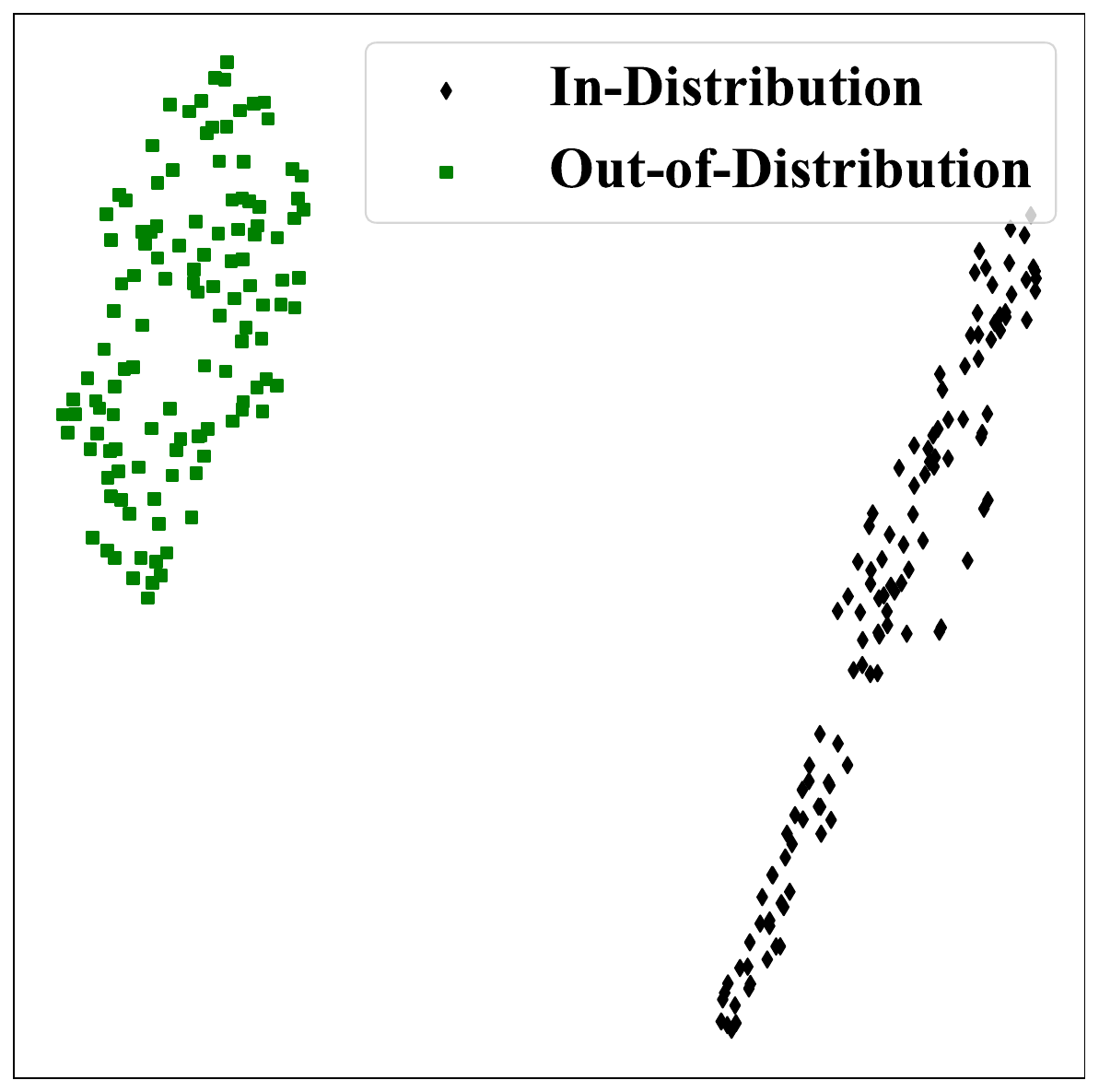}
  }
  \subfigure[\textbf{Evaluation} on MA Dataset  \newline @ \qquad Classifier Subspace \newline @ \qquad Accuracy: 0.65 ± 0.14]{
  \includegraphics[width=1.55in]{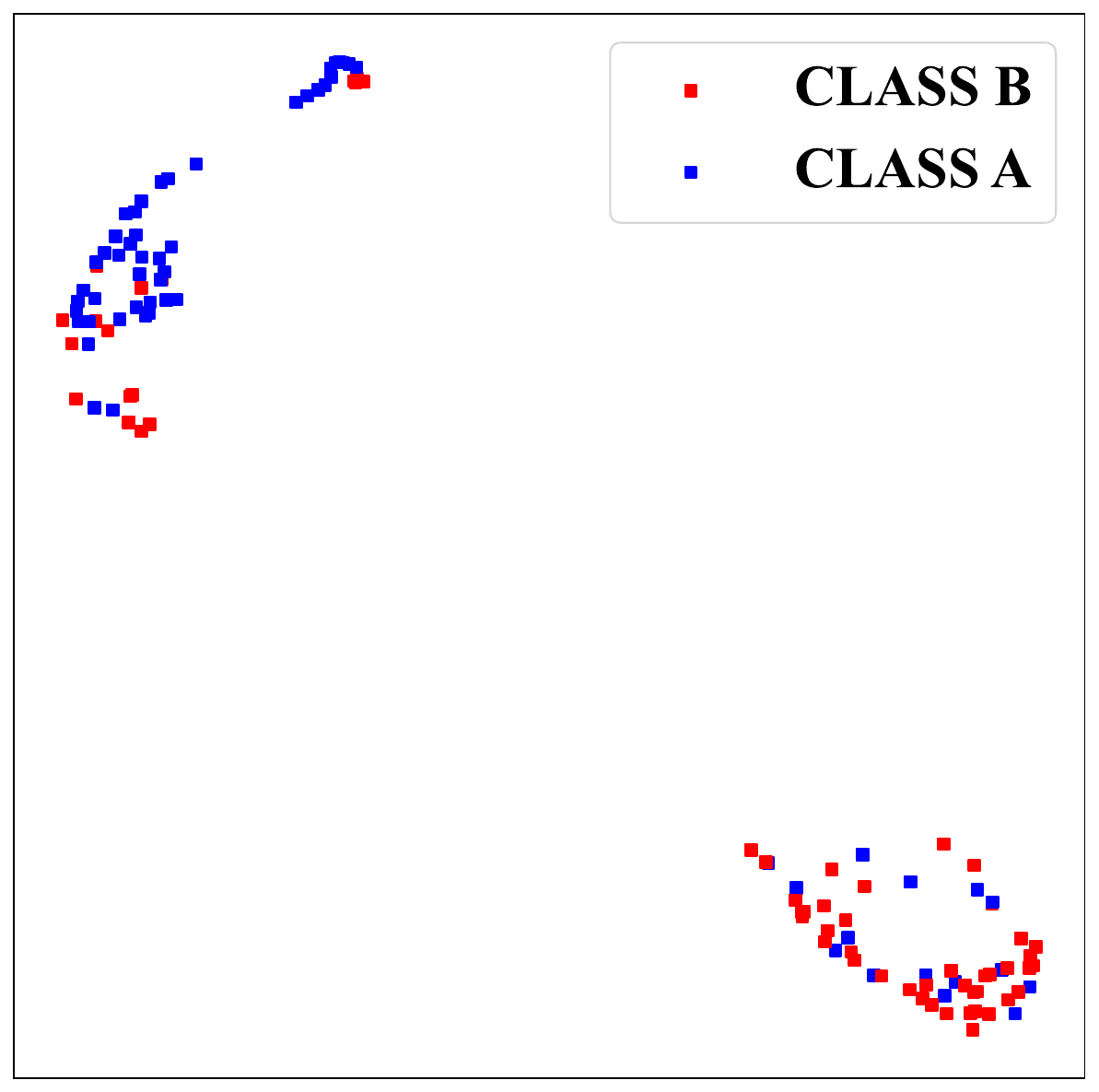}
  }
  \subfigure[\textbf{Evaluation} on UFFT Dataset \newline @ \qquad Classifier Subspace \newline @ \qquad Accuracy: 0.58 ± 0.19]{
  \includegraphics[width=1.55in]{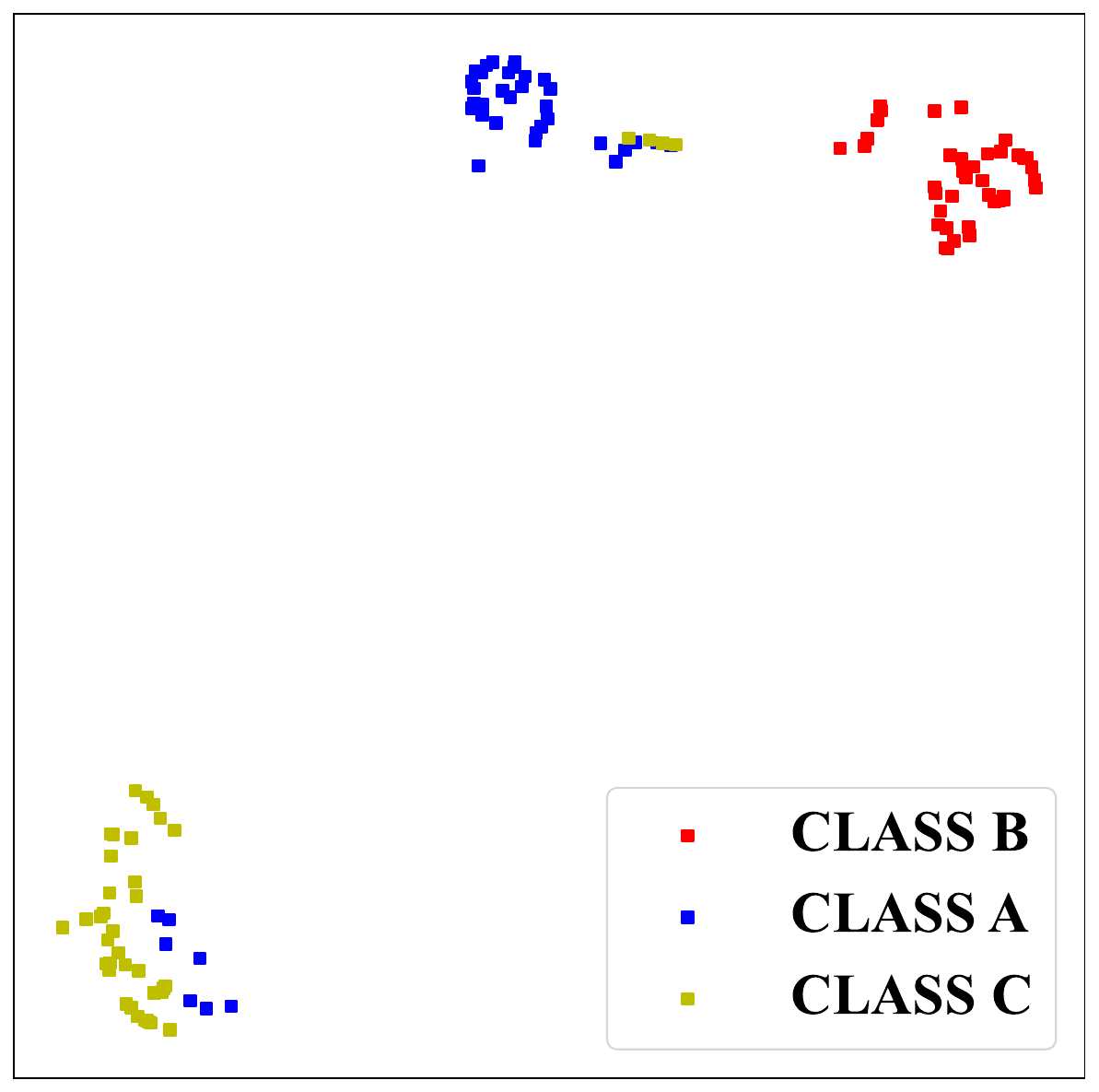}
  }
  \vspace{-0.1cm}
  \caption{Visualization of the subspaces of detector and classifier}
  \label{fig:class_subspace}
  \vspace{-0.1cm}
\end{figure}

We illustrate the feature vectors of the transformer-based neural network model, which excels in identifying and excluding OOD data in fNIRS, within both detection and classification subspaces as depicted in Figure \ref{fig:class_subspace}. 

This visualization confirms our experimental findings. Consistent with previous observations, a larger distance between feature vectors of different categories in the subspace indicates a clearer distinction. The visualization of the subspaces reveals the model's capability to separate feature vectors in both detection and classification subspaces, ensuring effective differentiation of categories within each subspace. Our approach maintains the classification performance while significantly increasing the separation between in-distribution and OOD data in the detection subspace.

\subsection{Shortcoming}

\begin{figure}[htpb]
  \centering
  \subfigure[\textbf{Evaluation} on UFFT Dataset \newline @ \qquad Detector Subspace]{
  \includegraphics[width=1.55in]{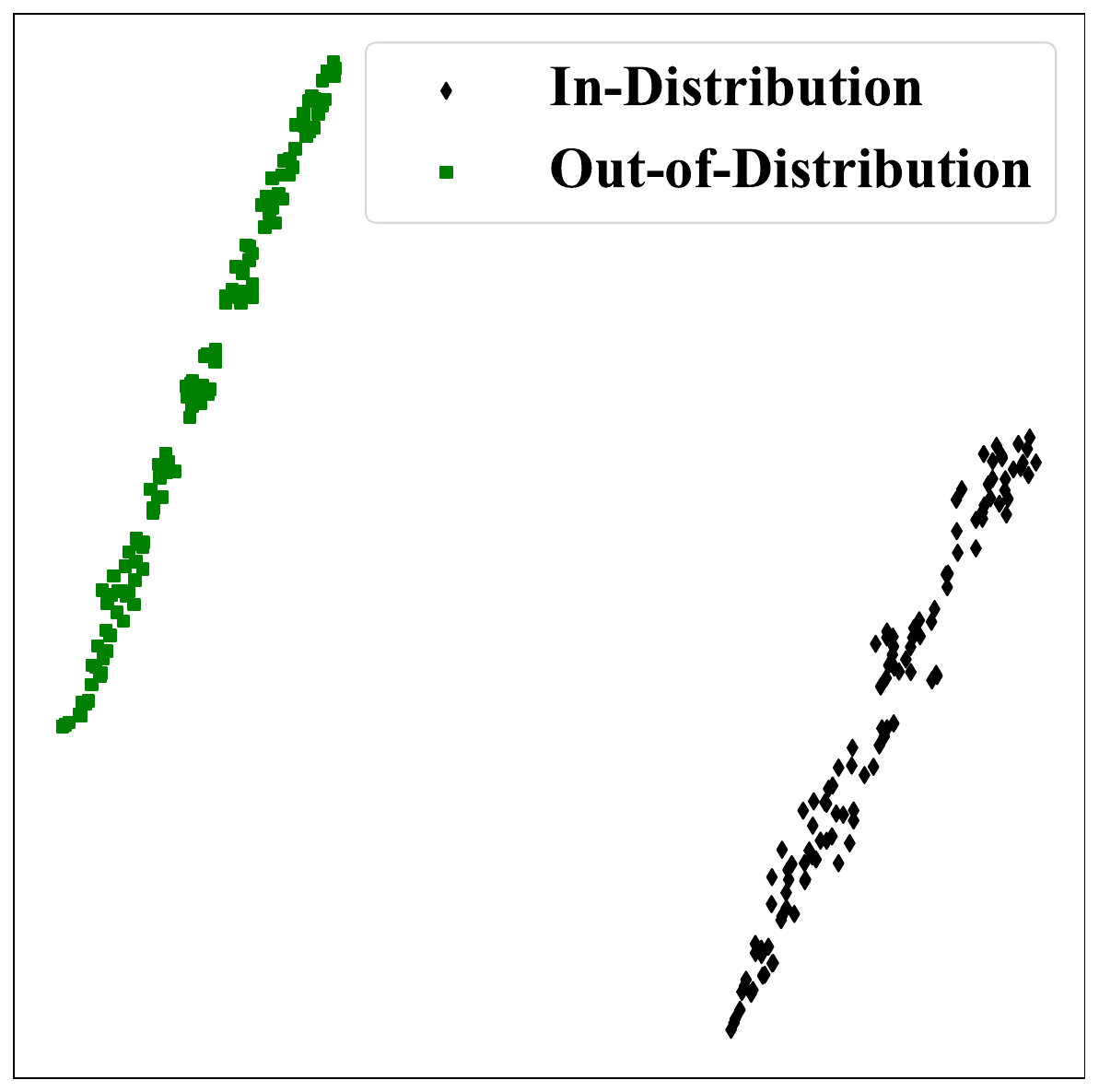}
  }
  \subfigure[\textbf{Evaluation} on UFFT Dataset \newline @ \qquad Classifier Subspace]{
  \includegraphics[width=1.55in]{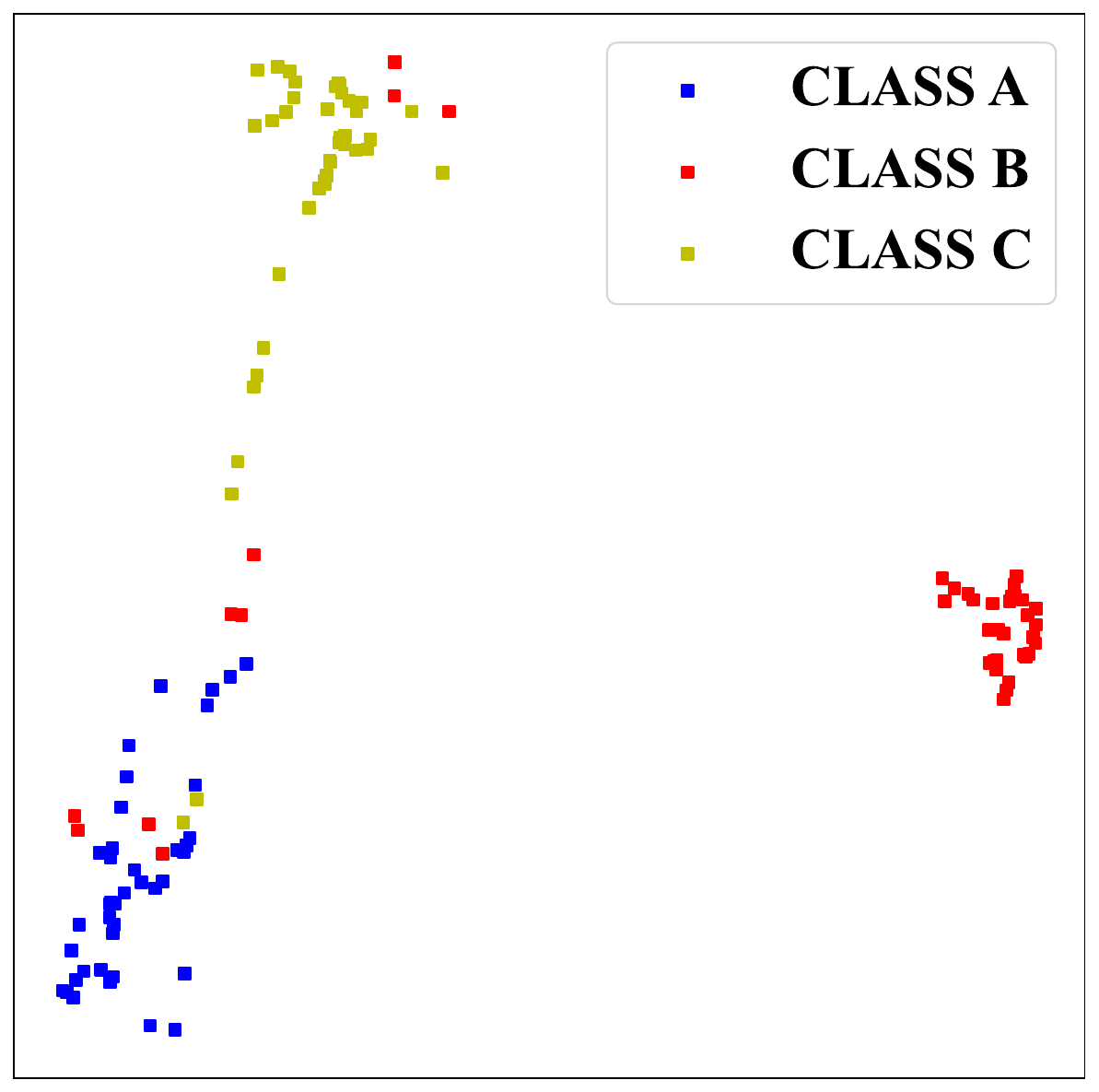}
  }
  \vspace{-0.1cm}
  \caption{Visualization of ill case in the subspaces of detector and classifier}
  \label{fig:ill_subspace}
  \vspace{-0.1cm}
\end{figure}

In our experiments, we also found that despite the network's design enabling the decoupling of detection and classification subspaces, we observed instances of coupling. As illustrated in Figure \ref{fig:ill_subspace}, while the network successfully separates feature vectors in the detection subspace, it struggles to do the same in the classification subspace. 

We theorize this occurs because metric loss drives a greater separation between in-distribution and OOD data in feature space, and concurrently, narrows the gap between different categories within the distribution. Consequently, despite the division into two subspaces at the Linear layer, the losses from both subspaces influence the preceding feature extraction layer. A more effective strategy would involve increasing the separation between different in-distribution categories within the detection subspace as 

\begin{equation}
\begin{aligned}
L &= \mathbb{E}[-metric(net(\hat{x}_{in})[:m][:n-1], net(\hat{x}_{out})[:n-1] \\
   & + \mathbb{E}[-metric(net(\hat{x}_{1_{in}})[:m][n:], net(\hat{x}_{2_{in}})[:m][n:])]
\end{aligned}
\end{equation}
where $net(\cdot)[:m][n:]$ is the first n-dimensional feature vector in the m-dimensional detection space, using to distinguish different classes of in-distribution data.

This approach of subspace nesting not only aids in excluding OOD data but also enhances the classification of in-distribution data, and we will implement it in the future.

\section{CONCLUSION \label{sec:conclusion}}

In this letter, we present our findings from experiments testing current fNIRS deep learning models, which tend to be overly confident when encountering OOD data. To address this, we introduce a two-stage training approach combining metric learning with supervision techniques, aimed at bolstering the models' capability to dismiss OOD data. Our results demonstrate the effectiveness of this method across various fNIRS models, particularly with transformer-based architectures.

\addtolength{\textheight}{-12cm}

\bibliography{references}

\begin{thebibliography}{10}

\bibitem{naseer2015fnirs}
Noman Naseer and Keum-Shik Hong.
\newblock fnirs-based brain-computer interfaces: a review.
\newblock {\em Frontiers in human neuroscience}, 9:3, 2015.

\bibitem{jobsis1977noninvasive}
Frans~F J{\"o}bsis.
\newblock Noninvasive, infrared monitoring of cerebral and myocardial oxygen sufficiency and circulatory parameters.
\newblock {\em Science}, 198(4323):1264--1267, 1977.

\bibitem{eastmond2022deep}
Condell Eastmond, Aseem Subedi, Suvranu De, and Xavier Intes.
\newblock Deep learning in fnirs: a review.
\newblock {\em Neurophotonics}, 9(4):041411--041411, 2022.

\bibitem{ferrari2012brief}
Marco Ferrari and Valentina Quaresima.
\newblock A brief review on the history of human functional near-infrared spectroscopy (fnirs) development and fields of application.
\newblock {\em Neuroimage}, 63(2):921--935, 2012.

\bibitem{buzsaki2012origin}
Gy{\"o}rgy Buzs{\'a}ki, Costas~A Anastassiou, and Christof Koch.
\newblock The origin of extracellular fields and currents—eeg, ecog, lfp and spikes.
\newblock {\em Nature reviews neuroscience}, 13(6):407--420, 2012.

\bibitem{logothetis2008we}
Nikos~K Logothetis.
\newblock What we can do and what we cannot do with fmri.
\newblock {\em Nature}, 453(7197):869--878, 2008.

\bibitem{shin2016open}
Jaeyoung Shin, Alexander von L{\"u}hmann, Benjamin Blankertz, Do-Won Kim, Jichai Jeong, Han-Jeong Hwang, and Klaus-Robert M{\"u}ller.
\newblock Open access dataset for eeg+ nirs single-trial classification.
\newblock {\em IEEE Transactions on Neural Systems and Rehabilitation Engineering}, 25(10):1735--1745, 2016.

\bibitem{chen2020classification}
Lei Chen, Qiang Li, Hong Song, Ruiqi Gao, Jian Yang, Wentian Dong, and Weimin Dang.
\newblock Classification of schizophrenia using general linear model and support vector machine via fnirs.
\newblock {\em Physical and Engineering Sciences in Medicine}, 43:1151--1160, 2020.

\bibitem{lyu2021domain}
Boyang Lyu, Thao Pham, Giles Blaney, Zachary Haga, Angelo Sassaroli, Sergio Fantini, and Shuchin Aeron.
\newblock Domain adaptation for robust workload level alignment between sessions and subjects using fnirs.
\newblock {\em Journal of Biomedical Optics}, 26(2):022908--022908, 2021.

\bibitem{sun2020novel}
Zhe Sun, Zihao Huang, Feng Duan, and Yu~Liu.
\newblock A novel multimodal approach for hybrid brain--computer interface.
\newblock {\em IEEE Access}, 8:89909--89918, 2020.

\bibitem{rojas2021pain}
Raul~Fernandez Rojas, Julio Romero, Jehu Lopez-Aparicio, and Keng-Liang Ou.
\newblock Pain assessment based on fnirs using bi-lstm rnns.
\newblock In {\em 2021 10th International IEEE/EMBS Conference on Neural Engineering (NER)}, pages 399--402. IEEE, 2021.

\bibitem{wang2022transformer}
Zenghui Wang, Jun Zhang, Xiaochu Zhang, Peng Chen, and Bing Wang.
\newblock Transformer model for functional near-infrared spectroscopy classification.
\newblock {\em IEEE Journal of Biomedical and Health Informatics}, 26(6):2559--2569, 2022.

\bibitem{wang2023rethinking}
Zenghui Wang, Jihong Fang, and Jun Zhang.
\newblock Rethinking delayed hemodynamic responses for fnirs classification.
\newblock {\em IEEE Transactions on Neural Systems and Rehabilitation Engineering}, 2023.

\bibitem{cao2024calibration}
Zhihao Cao and Zizhou Luo.
\newblock Calibration of deep learning classification models in fnirs.
\newblock {\em arXiv preprint arXiv:2402.15266}, 2024.

\bibitem{hendrycks2016baseline}
Dan Hendrycks and Kevin Gimpel.
\newblock A baseline for detecting misclassified and out-of-distribution examples in neural networks.
\newblock {\em arXiv preprint arXiv:1610.02136}, 2016.

\bibitem{liang2017enhancing}
Shiyu Liang, Yixuan Li, and Rayadurgam Srikant.
\newblock Enhancing the reliability of out-of-distribution image detection in neural networks.
\newblock {\em arXiv preprint arXiv:1706.02690}, 2017.

\bibitem{lee2017training}
Kimin Lee, Honglak Lee, Kibok Lee, and Jinwoo Shin.
\newblock Training confidence-calibrated classifiers for detecting out-of-distribution samples.
\newblock {\em arXiv preprint arXiv:1711.09325}, 2017.

\bibitem{masana2018metric}
Marc Masana, Idoia Ruiz, Joan Serrat, Joost van~de Weijer, and Antonio~M Lopez.
\newblock Metric learning for novelty and anomaly detection.
\newblock {\em arXiv preprint arXiv:1808.05492}, 2018.

\bibitem{mohseni2020self}
Sina Mohseni, Mandar Pitale, JBS Yadawa, and Zhangyang Wang.
\newblock Self-supervised learning for generalizable out-of-distribution detection.
\newblock In {\em Proceedings of the AAAI Conference on Artificial Intelligence}, volume~34, pages 5216--5223, 2020.

\bibitem{bak2019open}
SuJin Bak, Jinwoo Park, Jaeyoung Shin, and Jichai Jeong.
\newblock Open-access fnirs dataset for classification of unilateral finger-and foot-tapping.
\newblock {\em Electronics}, 8(12):1486, 2019.

\bibitem{cope1988system}
Mark Cope and David~T Delpy.
\newblock System for long-term measurement of cerebral blood and tissue oxygenation on newborn infants by near infra-red transillumination.
\newblock {\em Medical and Biological Engineering and Computing}, 26:289--294, 1988.

\bibitem{lamer2022standardized}
Antoine Lamer, Mathilde Fruchart, Nicolas Paris, Benjamin Popoff, Ana{\"\i}s Payen, Thibaut Balcaen, William Gacquer, Guillaume Bouzill{\'e}, Marc Cuggia, Matthieu Doutreligne, et~al.
\newblock Standardized description of the feature extraction process to transform raw data into meaningful information for enhancing data reuse: Consensus study.
\newblock {\em JMIR Medical Informatics}, 10(10):e38936, 2022.

\end{thebibliography}
\bibliographystyle{unsrt}

\end{document}